\documentclass[aps,pra,preprint,superscriptaddress,longbibliography]{revtex4-2}

\usepackage{amsmath}
\usepackage{amsthm}
\usepackage{amsfonts}
\usepackage{amssymb}
\usepackage{xcolor,graphicx}
\usepackage{tikz}
\usepackage{color}
\usepackage[pdftex]{hyperref} 
\usepackage{longtable}
\usepackage{array}
\usepackage{dsfont}

\begin{document}
	
	\title{Non-local scattering control in coupled resonator networks}
	
	\author{D. A. Rom\'an-Garc\'ia }
	\email[e-mail: ]{a01636212@tec.mx}
	\affiliation{Tecnologico de Monterrey, Escuela de Ingenier\'ia y Ciencias, Ave. Eugenio Garza Sada 2501, Monterrey, N.L., Mexico, 64849}	
		
	\author{F. H. Maldonado-Villamizar}
	\email[e-mail: ]{fmaldonado@inaoep.mx}
	\affiliation{CONACYT - Instituto Nacional de Astrof\'isica, \'Optica y Electr\'onica, Calle Luis Enrique Erro No. 1. Sta. Ma. Tonantzintla, Pue. C.P. 72840, Mexico}
	
	\author{B. Jaramillo-\'Avila}
	\email[e-mail: ]{jaramillo@inaoep.mx}
	\affiliation{CONACYT - Instituto Nacional de Astrof\'isica, \'Optica y Electr\'onica, Calle Luis Enrique Erro No. 1. Sta. Ma. Tonantzintla, Pue. C.P. 72840, Mexico}

	\author{B. M. Rodr\'iguez-Lara}
	\email[e-mail: ]{bmlara@tec.mx}
	\affiliation{Tecnologico de Monterrey, Escuela de Ingenier\'ia y Ciencias, Ave. Eugenio Garza Sada 2501, Monterrey, N.L., Mexico, 64849}	
	
	\date{\today}
	
	\begin{abstract}
	We demonstrate scattering control of Gaussian-like wave packets propagating with constant envelope velocity and invariant waist through coupled resonator optical waveguides (CROW) via an external resonator coupled to multiple sites of the CROW. 
	We calculate the analytical reflectance and transmittance using standard scattering methods from waveguide quantum electrodynamics and show it is possible to approximate them for an external resonator detuned to the CROW.
	Our analytical and approximate results are in good agreement with numerical simulations.
	We engineer various configurations using an external resonator coupled to two sites of a CROW to show light trapping with effective exponential decay between the coupling sites, wave packet splitting into two pairs of identical Gaussian-like wave packets, and a non-local Mach-Zehnder interferometer.
	\end{abstract}
	
	
	\maketitle
	\newpage
\section{Introduction}

Coupled resonator optical waveguides (CROW) are a versatile platform for integrated photonic devices \cite{Morichetti2012}.
They may be fabricated to function in linear or nonlinear regimes \cite{Melloni2003} under different coupling topologies \cite{Tsay2011} that strongly influence their spectral behavior; for example, allowing group velocity modulation of optical wave packets \cite{Khurgin2008}. 
They are particularly promising for applications that require phase modulators \cite{Taylor1999,Shaw1999}, light filtering \cite{Madsen2000}, second-harmonic generation \cite{Xu2000}, slow light \cite{Morichetti2012}, light trapping \cite{Poon2004,Scheuer2005}, and non-linear signal processing \cite{Kazanskiy2014}, for example, such as high-capacity telecommunication networks \cite{Kumar2020,Kumar2021} or quantum information processing \cite{Takesue2013}. 
They consist of weakly coupled, independent, high-Q optical resonators \cite{Yariv1999} that may take the form of microrings \cite{Chremmos2010,Heebner2007,Rabus2007,Bogaerts2005,Xia2007}, microdisks \cite{Hu2008,Hu2010}, microspheres \cite{Vahala2003,Kapitonov2007}, microtoroids \cite{Kippenberg2007,Schliesser2008,Hossein2010}, square microresonators \cite{Mookherjea2008}, Fabry-P\'erot cavities \cite{Goeckeritz2010}, or photonic crystal cavities \cite{Akahane2003,Karle2002}. 
Many of these may be integrated into photonic devices. 
However, microrings are perhaps more widely used due to the high level of control and reproducibility in their fabrication \cite{Morichetti2012}. 

CROWs are also a viable platform for photonic simulation of light-matter interactions \cite{Kockum2019a}.
An infinite chain of identical and uniformly coupled resonators supports a continuum of Bloch states, playing the role of boson field modes, that may be coupled to additional external resonators, playing the role of artificial atoms, to simulate interesting light-matter interactions from quantum electrodynamics (QED) that may not be readily available in current experiments.
For example, a single external resonator coupled to a single site of the CROW allows for the photonic simulation of the Purcell effect \cite{Vahala2003}, that is, spontaneous emission rate enhancement by the continuum.
The interaction of so-called giant atoms with electromagnetic modes in waveguide-QED is an example of more complex interactions suitable for photonic simulation in CROW-based devices.
Here, artificial atoms with size comparable to, or larger than, the wavelength of the electromagnetic \cite{Andersson2019,Wen2019,Guo2020} or acoustic \cite{Ask2019,Delsing2019} boson field mode, non-locally interact with them to produce interference effects unavailable to standard experiments in the long wavelength approximation.
For example, it is possible to optically simulate frequency-dependent coupling between a giant atom and the continuum \cite{Kockum2014}, decoherence-free interactions between two or more giant atoms mediated by the continuum \cite{Kockum2018}, suppression and enhancement of an excited giant atom decaying into a continuum reservoir \cite{Longhi2020,Jaramillo2020}, or single-photon scattering by a giant atom \cite{Zhao2020}. 

Here, we focus on studying Gaussian-like wave packets propagating with constant velocity and invariant waist along CROWs composed by identical and uniformly coupled resonators. 
These wave packets interact with an external resonator coupled to one or two sites on the CROW producing new reflected and transmitted wave packets that conserve the initial Gaussian waist. 
In Section \ref{section:model}, we introduce our model and review the propagation of Gaussian wave packets in a CROW.
Next, we discuss the scattering of these wave packets when a single external resonator couples to one site in the CROW, Sec. \ref{section:local}; we show that the reflection and transmission coefficients are controlled by two parameters, the ratios of the coupling to the external resonator and the detuning between the resonant frequencies of the external and CROW resonators with respect to the inter-site coupling in the CROW. 
We calculate the reflectance and transmittance for this system and show a Mach-Zehnder interferometer that uses the endpoints of a finite CROW as mirrors. 
In Section \ref{section:nonlocal}, we discuss wave packet scattering with a so-called giant atom, that is, when the external resonator couples non-locally to the CROW at different sites. 
We explore various regimes coupling the external resonator to two individual sites in the CROW.
First, we show effective light trapping with exponential excitation decay in the section of the CROW bounded by the sites that couple to the external resonator. 
Then, we demonstrate a wave packet splitter when the size of this section is comparable to the waist of the Gaussian wave packet. 
Lastly, we show a non-local Mach-Zehnder interferometer for a finite CROW with four interfering paths.
We close with our conclusions in Sec. \ref{section:conclusions}.

\section{Invariant wave packet propagation in a CROW}\label{section:model}

\begin{figure*}
	\centering
	\includegraphics{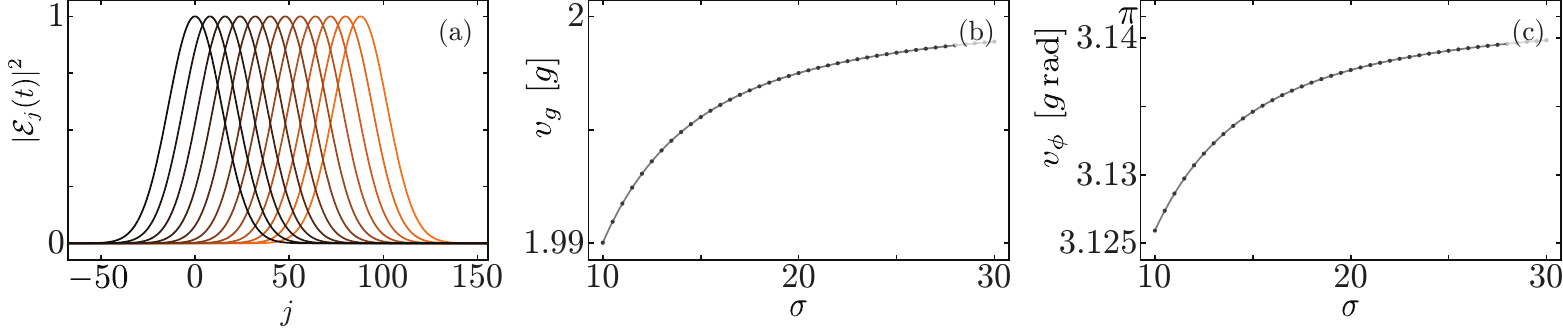}
	\caption{(a) Time evolution of an initial Gaussian-like complex field mode amplitude configuration with waist $\sigma=20$, initial phase $\theta=\pi/2$, and centered around the $j_{0}= 0$ resonator at scaled evolution times $gt=0,4,8,\ldots,36,40,44$ (black to orange). 
	(b) Group and (c) phase velocity as a function of the Gaussian-like wave packet waist $\sigma$ for initial phase $\theta = \pi/2$. The dots correspond to numerical experiments and the solid lines to the analytical expressions in Eq. (\ref{eq:speed}) and (\ref{eq:phasespeed}), in that order. 
	Numerical and analytical results show an absolute difference of order $10^{-5}$ or less.}\label{fig:Fig1}
\end{figure*}

Our first goal is to create insight on invariant wave packet propagation in a CROW. 
We use coupled mode theory to describe an ideal CROW composed of identical, single-mode resonators \cite{Jaramillo2020}, 
\begin{align}
	 i \partial_{t} \mathcal{E}_{j}(t) = g \left[ \mathcal{E}_{j-1}(t) + \mathcal{E}_{j+1}(t) \right],
\end{align}
where the index $j$ labels the resonator position in the CROW, the complex time-dependent function $\mathcal{E}_{j}(t)$ is the dimensionless amplitude of the localized field mode at the $j$-th resonator, and the real constant $g$ is the coupling strength between pairs of adjacent resonators that we will refer to as inter-site coupling strength from now on. 

An initial Gaussian-like complex field mode amplitude configuration \cite{Mookherjea2002a,Longhi2015}, 
\begin{align} \label{eq:initialprof}
	\mathcal{E}_{j}(0) = e^{-\frac{(j-j_{0})^2}{2\sigma^{2}}} e^{-i (j-j_{0})\theta},
\end{align}
centered around the $j_{0}$-th resonator with Gaussian waist $\sigma$, will propagate maintaining its shape and relative phase to the right (left) for the specific initial phase difference value $\theta = \pi/2 ~(-\pi/2)$, Fig. \ref{fig:Fig1}(a).
The initial wave packet envelope will travel with velocity, 
\begin{align}\label{eq:speed}
	v_{g} = 2g~e^{-\frac{1}{2\sigma^{2}}} ~ \sin \theta,
\end{align}
while its phase propagates with velocity,
\begin{align}\label{eq:phasespeed}
	v_{\phi} = 2g~e^{-\frac{1}{2\sigma^{2}}} \left( \cos \theta + \theta \sin \theta \right),
\end{align}
as shown in Fig. \ref{fig:Fig1}(b) and Fig. \ref{fig:Fig1}(c), in that order.
Due to its invariant propagation, this will be our initial field amplitude configuration to show the control of local and non-local scattering processes.

\section{Local scattering control}\label{section:local}

\begin{figure*}
	\centering
	\includegraphics{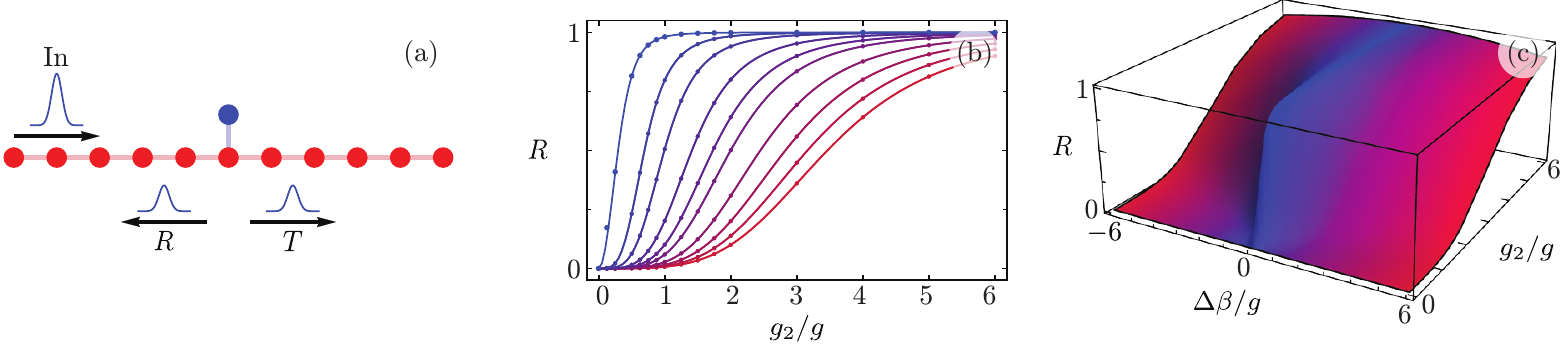}
	\caption{(a) Scattering configuration for an external resonator coupled to a single element in a CROW. 
	Panels (b) and (c) display the reflectance as a function of the normalized single-mode resonant frequency detuning $\Delta\beta / g$ and normalized external resonator coupling strength $g_{2}/g$. 
	The leftmost blue curve or region corresponds to $\Delta\beta / g = 0$ and the rightmost red to $\Delta\beta / g = \pm 6$.}\label{fig:Fig2}
\end{figure*}

Adding an external resonator, coupled to just the $k$-th site in the CROW, provides us with a coupled mode theory model, 
\begin{equation} \label{eq:differenceq}
\begin{aligned}
	i \partial_{t} \mathcal{E}_{j}(t) =& g \left[ \mathcal{E}_{j-1}(t) + \mathcal{E}_{j+1}(t) \right] + g_{2} \delta_{j,k} \mathcal{E}_{e}(t), \\
	i \partial_{t} \mathcal{E}_{e}(t) =& \Delta\beta \mathcal{E}_{e}(t) + g_{2} \mathcal{E}_{k}(t),
\end{aligned}
\end{equation}
where the real valued parameters $\Delta \beta$ and $g_{2}$ are the single-mode resonant frequency detuning between the external resonator and CROW, and the coupling strength between the external resonator and the $k$-th component in the CROW, in that order.
This model may be drawn into an analogy to single-photon scattering by a point-like atom coupled to an optical waveguide \cite{Jaramillo2020}. 
We calculate the scattering of a Gaussian-like wave packet, Eq. (\ref{eq:initialprof}), using standard scattering techniques that yield reflectance and transmittance coefficients, Appendix \ref{appendix:scattering}. 
For a single-mode resonant frequency detuning far from zero, we are ablo to approximate these coefficients,
\begin{equation}
	\begin{aligned}\label{eq:rtanalytic}
		R = \left| \frac{g_{2}^{2}}{g_{2}^{2}+2 i g \Delta \beta} \right|^2, 
		\qquad
		T = \left| \frac{2 g \Delta \beta}{g_{2}^{2}+2 i g \Delta \beta} \right|^2, \qquad \mathrm{for} ~ |\Delta\beta| \gtrsim 0.25,
	\end{aligned}
\end{equation}
considering an initial wave packet configuration placed to the left of the coupling point between the external resonator and CROW. 
The initial wave packet propagates to the right and we call reflectance (transmittance) the fraction of the total squared field amplitudes that propagates to the left (right), 
\begin{align}
	R = \frac
		{ \sum_{j = -\infty}^{k} \vert \mathcal{E}_{j}(t_{0}) \vert^{2} }
		{ \sum_{j = - \infty}^{\infty} \vert \mathcal{E}_{j}(0) \vert^{2} }, \qquad 
	T = \frac
		{ \sum_{j = k+1}^{\infty} \vert \mathcal{E}_{j}(t_{0}) \vert^{2} }
		{ \sum_{j = - \infty}^{\infty} \vert \mathcal{E}_{j}(0) \vert^{2} }, 
\end{align}
after interaction with the external resonator coupled to the $k$-th resonator in the CROW at a latter time $t_{0} \gg 0$, Fig. 2(a). 
Figure \ref{fig:Fig2} displays both the analytical coefficient $R$ and its value from a numerical simulation. 
The simulation has incoming and outgoing wave packets of waist $\sigma = 20$ and the CROW has $501$ components with the external resonator coupled to the $251$-th element. 
Figure \ref{fig:Fig2}(b) presents a comparison between theory (solid lines) and numerical results (dots) for the reflectance. It considers different resonant frequency detunings $\Delta \beta$ and external coupling strengths $g_{2}$ in terms of the inter-site coupling strength $g$. 
The line at zero resonant frequency detuning $\Delta \beta = 0$, where the approximation for the analytical results in Eq. (\ref{eq:rtanalytic}) breaks down, is calculated using the full expressions in Appendix \ref{appendix:scattering}.
Figure \ref{fig:Fig2}(c) shows an interpolation of the numerical results in 3D using the same color coding than Fig. \ref{fig:Fig2}(b). 

\subsection{Mach-Zehnder interferometer}\label{subsection:onesiteinterferometer}

\begin{figure*}
	\centering
	\includegraphics[width=0.8\textwidth]{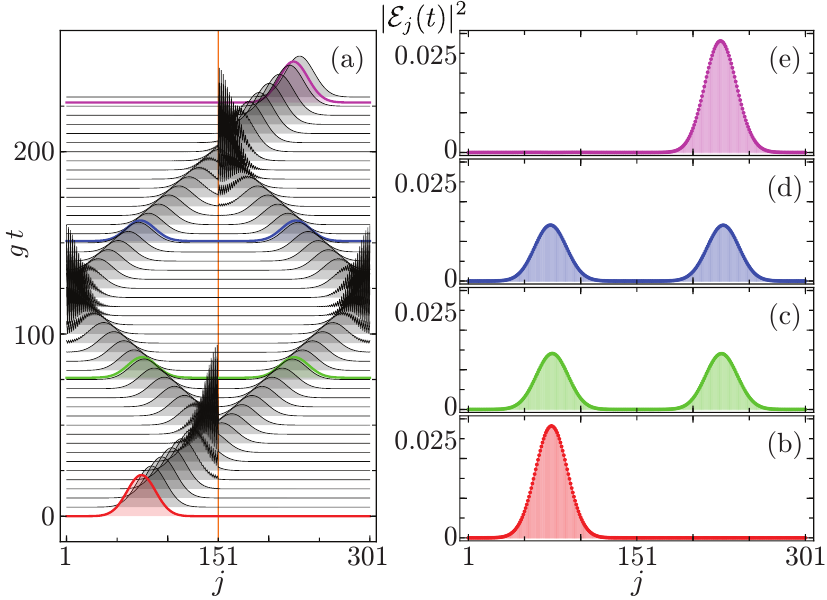}
	\caption{
		(a) Squared field amplitude, $\vert \mathcal{E}_{j}(t)\vert^{2}$, at each resonator, $j$, for different scaled evolution times $gt$ in an effective Mach-Zehnder interferometer. 
		The right column shows the same squared field amplitude for scaled evolution times 
		(b) before interaction with the external resonator, $g t = 0$; 
		(c) after interaction and before reflection at the endpoints of the finite CROW, $g t = 76$; 
		(d) after reflection and before a second interaction with the external resonator, $g t = 151$; and 
		(e) after the second interaction with the external resonator, $g t = 227$, where the original wave packet envelope is approximately reconstructed. 
		All wave packets have waist $\sigma = 20$.}\label{fig:Fig3}
\end{figure*}

We use the insight from previous results, and the fact that the endpoints of a finite CROW act as perfect mirrors for our wave packets, to construct a Mach-Zehnder interferometer \cite{Martinez2003}, Fig. \ref{fig:Fig3}. 
The configuration is identical to that in Fig. \ref{fig:Fig2}(a) but with a finite CROW composed by $301$ identical resonators. 
We couple an external resonator to the $151$-th element in the CROW 
and set both the resonant frequency detuning and external coupling equal to twice the inter-site coupling strength, $\Delta \beta = g_{2} = 2 g$. 
This produces a 50/50 reflectance to transmittance ratio in the system. 
Our initial wave packet starts at the left of the external resonator and propagates to the right, Fig. \ref{fig:Fig3}(b). 
After the interaction, the reflected and transmitted wave packets propagate and reflect at the left and right endpoints of the CROW. 
Both wave packets travel identical path lengths to, Fig. \ref{fig:Fig3}(c), and from, Fig. \ref{fig:Fig3}(d) the endpoints of the finite CROW. 
They recombine at the external resonator and continue the initial trajectory, Fig. \ref{fig:Fig3}(e), as expected.
It is possible to observe the interference of the incoming wave packet and its reflection at the external resonator and endpoints of the finite CROW, Fig. \ref{fig:Fig3}(a), once the interaction begins.

\section{Non-local scattering control}\label{section:nonlocal}

\begin{figure*}
	\centering
	\includegraphics{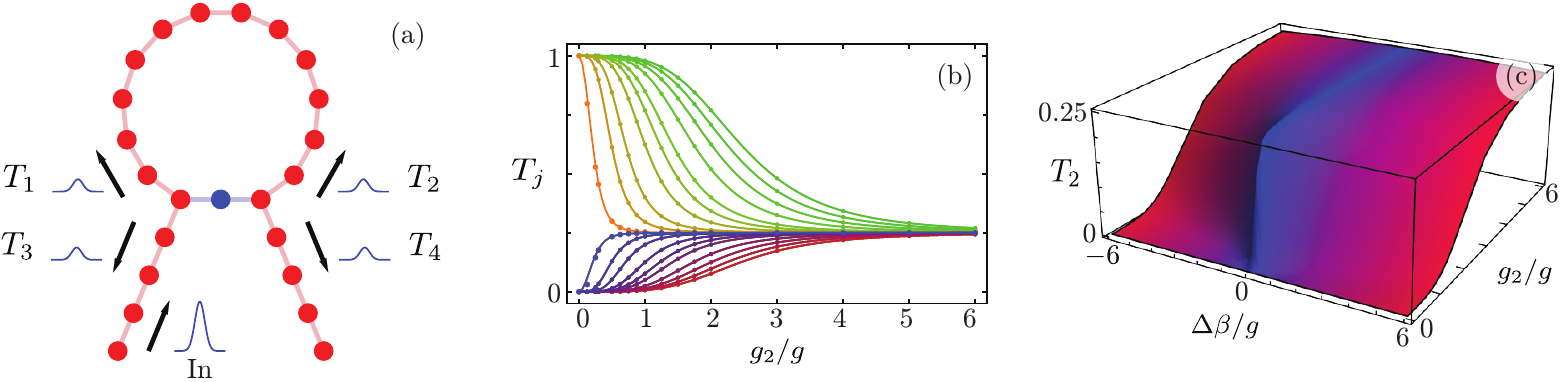}
	\caption{(a) Scattering configuration for an external resonator equally coupled to a pair of elements in a CROW. 
		(b) The transmission corresponding to the four wave packets $T_{1}$, $T_{2}$, $T_{3}$, $T_{4}$ as a function of the normalized single-mode frequency detuning $\Delta\beta / g$ and normalized coupling strength to the external resonator $g_{2}/g$. 
		The yellow to green curves show the first transmission $T_{1}$ and blue to red curves the other three, which are always equal, $T_{2}=T_{3}=T_{4}$.
		The leftmost yellow and blue curves correspond to $\Delta\beta / g= 0$ and the rightmost green and red curves correspond to $\Delta\beta / g= \pm 6$. 
		Panel (c) displays $T_{2}=T_{3}=T_{4}$ using the same color coding.}\label{fig:Fig4}
\end{figure*} 

So far, we discussed point-like interactions between external resonators and the CROW, that is, single site, local coupling. 
We now turn beyond point-like interactions and explore an external resonator coupled to several sites on the CROW. 
This provides a platform to simulate the interaction of giant atoms with the continuum \cite{Jaramillo2020}. 
We deal with the simplest possible case, just one external resonator coupled to two CROW sites. 
When an incoming Gaussian-like wave packet encounters the external resonator, four new wave packets are created, Fig. \ref{fig:Fig4}(a). 
Standard scattering techniques yield analytical transmittance coefficients, Appendix \ref{appendix:scattering}, feasible for approximation when the single-mode resonant frequency detuning is sufficiently far from zero, 
\begin{equation}\label{eq:rtanalytict1234}
\begin{aligned}
	T_{1} &= \left| \frac{g_{2}^{2}-2 i g \Delta \beta}{2 g_{2}^{2} - 2 i g \Delta \beta} \right|^{2}, \\
	T_{2} = T_{3} = T_{4} &= \left| \frac{g_{2}^{2}}{2 g_{2}^{2} - 2 i g \Delta \beta} \right|^{2}, 		
\end{aligned}
	\qquad \mathrm{for} ~ \vert \Delta \beta \vert \gtrsim 0.25.
\end{equation}
These are valid before wave packets one and two propagate and encounter themselves or the external resonator again. 
A numerical simulation yields these coefficients in the following manner,
\begin{equation}
\begin{aligned}
	T_{1} =& \frac
		{ \sum_{j = k_{1}+1}^{l}		\vert \mathcal{E}_{j}(t_{0}) \vert^{2} }
		{ \sum_{j = -\infty}^{\infty} 	\vert \mathcal{E}_{j}(0) \vert^{2} } ,
\qquad \qquad
	T_{2} =& \frac
		{ \sum_{j = l+1}^{k_{2}} 		\vert \mathcal{E}_{j}(t_{0}) \vert^{2} }
		{ \sum_{j = -\infty}^{\infty} 	\vert \mathcal{E}_{j}(0) \vert^{2} }, 
\\
	T_{3} =& \frac
		{ \sum_{j = -\infty}^{k_{1}} 	\vert \mathcal{E}_{j}(t_{0}) \vert^{2} }
		{ \sum_{j = -\infty}^{\infty} 	\vert \mathcal{E}_{j}(0) \vert^{2} } ,
\qquad \qquad
	T_{4} =& \frac
		{ \sum_{j = k_{2}+1}^{\infty} 	\vert \mathcal{E}_{j}(t_{0}) \vert^{2} }
		{ \sum_{j = -\infty}^{\infty} 	\vert \mathcal{E}_{j}(0) \vert^{2} },
\end{aligned}
\end{equation}
where the external resonator couples to the CROW at the $k_{1}$-th and $k_{2}$-th sites, the $l$-th site lies between the $k_{1}$-th and $k_{2}$-th sites, at any given times $t_{0}$ after the interaction but before wave packets one and two propagate and encounter each other or the external resonator again. 

Figure \ref{fig:Fig4} displays results for both analytical and numerical transmission coefficients. 
The simulation has incoming and outgoing wave packets of waist $\sigma = 20$ interacting with an external resonator coupled to the 241-th and 482-th sites of a CROW with 682 resonators. 
Figure \ref{fig:Fig4}(b) presents a comparison between analytical (solid lines) and numerical results (dots) for the four transmission coefficients. 
Again, the line at zero resonant frequency detuning $\Delta \beta = 0$, where the approximation for the analytical results in Eq. (\ref{eq:rtanalytict1234}) breaks down, is calculated using the full expressions in Appendix \ref{appendix:scattering}.

\subsection{Exponential decay of trapped light}

\begin{figure*}
	\centering
	\includegraphics[width=0.8\textwidth]{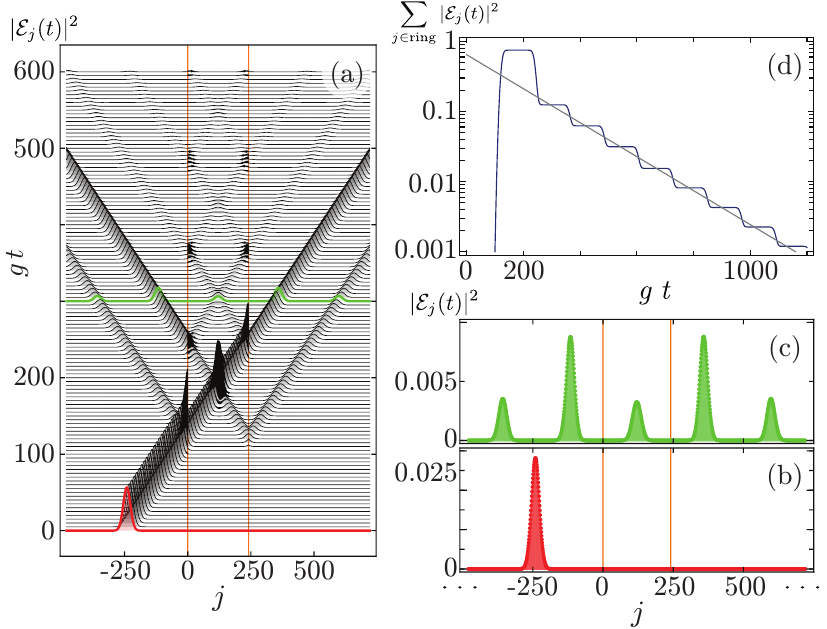}
	\caption{Gaussian-like wave packet scattering by an external resonator coupled to two sites of a CROW (orange vertical lines) creating the ring structure in Fig. \ref{fig:Fig4}(a). 
		(a) Squared field amplitude at the $j$-th resonator $\vert \mathcal{E}_{j}(t)\vert^{2}$ for different scaled evolution times: (b) $g t = 0$ and (c) $g t = 300$. 
		(d) Effective exponential decay of the total squared field amplitude inside the ring structure. 
		All wave packets have waist $\sigma = 20$.}\label{fig:Fig5}
\end{figure*}

We now focus on a large ring structure bounded between two sites coupled to an external resonator, Fig. \ref{fig:Fig4}(a). 
We use a numerical experiment to demonstrate quasi-exponential decay of the total squared field amplitude trapped in the ring structure, see Fig. \ref{fig:Fig5}. 
Here, an initial Gaussian-like wave packet propagates through a very large CROW, see Fig. \ref{fig:Fig5}(b). 
The external resonator couples to the CROW at sites $j=0$ and $j=241$, producing a ring structure with 240 sites. 
When the initial wave packet interacts with the external resonator, four new wave packets are created. 
Two propagate 
	inside the ring structure, $T_{1}$ and $T_{2}$, 
and two propagate 
	away from it, $T_{3}$ and $T_{4}$. 
Since the CROW is very large, the wave packets that propagate away from the ring structure do not interact with it again. 
As the four wave packets propagate, $T_{1}$, $T_{2}$, $T_{3}$, and $T_{4}$, the total squared field amplitude in the ring structure resonators remains constant, that is, light is trapped producing the first plateau in Fig. \ref{fig:Fig5}(d). 
Eventually the wave packets propagating inside the ring structure, $T_{1}$ and $T_{2}$, reach the external resonator again and get scattered, producing a sharp fall as a fraction of the total squared field amplitude leaves the ring structure. 
This generates new wave packets that propagate both away and inside the ring structure. 
Each interaction with the external resonator takes away some constant fraction of the total squared field amplitude from the ring structure, producing an effective exponential decay via plateaus and falls, Fig. \ref{fig:Fig5}(d). 
This quasi-exponential decay scenario occurs when the waist of the Gaussian-like wave packets is smaller than the size of the ring structure.
Otherwise, the propagation time inside the ring structure before reaching the external resonator is too short for the formation of plateaus.
Here,  Fig. \ref{fig:Fig5}, we choose the single-mode resonant frequency detuning and the coupling to the external resonator equal to the inter-site coupling strength, $\Delta \beta = g_{2}= g$, to produce transmittances $T_{1}=0.625$ and $T_{2}=T_{3}=T_{4}=0.125$, such that most of the field propagates inside the ring structure; only a quarter of the total squared field amplitude propagates away from it. 
%

\subsection{Wave packet splitter}

\begin{figure*}
	\centering
	\includegraphics[width=0.8\textwidth]{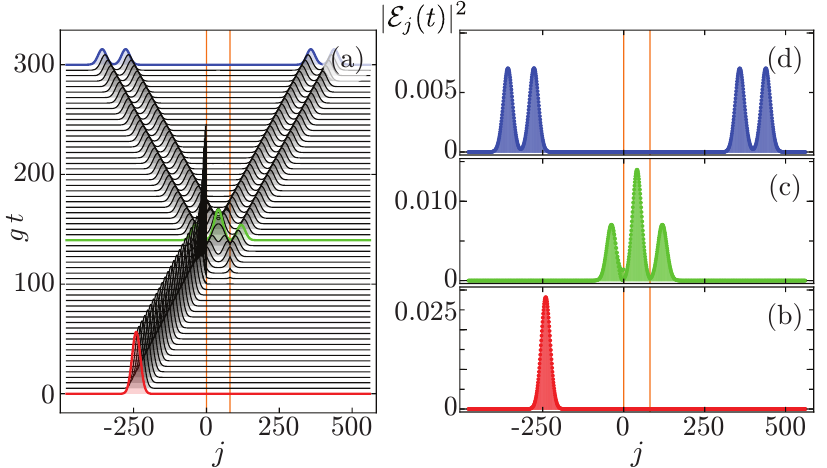}
	\caption{Double wave packet generation from the interaction between an external resonator and two sites of an effective infinite CROW (orange vertical lines) creating the ring structure in Fig. \ref{fig:Fig4}(a). 
		(a) Squared field amplitude at the $j$-th resonator $\vert \mathcal{E}_{j}(t)\vert^{2}$ for different scaled evolution times: 
		(b) $g t = 0$, when the original wave packet is propagating to the right. 
		(c) $g t = 140$, when the wave packet is interacting with the small ring structure. 
		(d) $g t = 300$, with two pairs of identical wave packets propagating to the left and right. 
		All wave packets have waist $\sigma = 20$.}\label{fig:Fig6}
\end{figure*}

Figure \ref{fig:Fig6}(a) shows the squared field amplitude propagation for a ring structure with size comparable to the Gaussian-like wave packet waist.
We consider an external resonator coupled to a CROW at two sites, $j=0$ and $j=81$, with zero single-mode resonant frequency detuning $\Delta \beta = 0$ and coupling strength equal to twice the inter-site coupling strength $g_{2} = 2 g$ leading to equal transmissions $T_{1}=T_{2}=T_{3}=T_{4}=0.25$.
We engineered this configuration to take an initial wave packet with waist $\sigma = 20$, Fig. \ref{fig:Fig6}(b), that, upon interaction with the external resonator, Fig. \ref{fig:Fig6}(c), produces two pairs of identical wave packets, with the same waist than the original, propagating away from the ring structure, Fig. \ref{fig:Fig6}(d).
One pair propagates to the left, and the other pair to the right, \ref{fig:Fig6}(a).
We optimize the ring structure size to capture a negligible fraction of the total field amplitude in it.

\subsection{Mach-Zehnder interferometer}

\begin{figure*}
	\centering
	\includegraphics[width=0.8\textwidth]{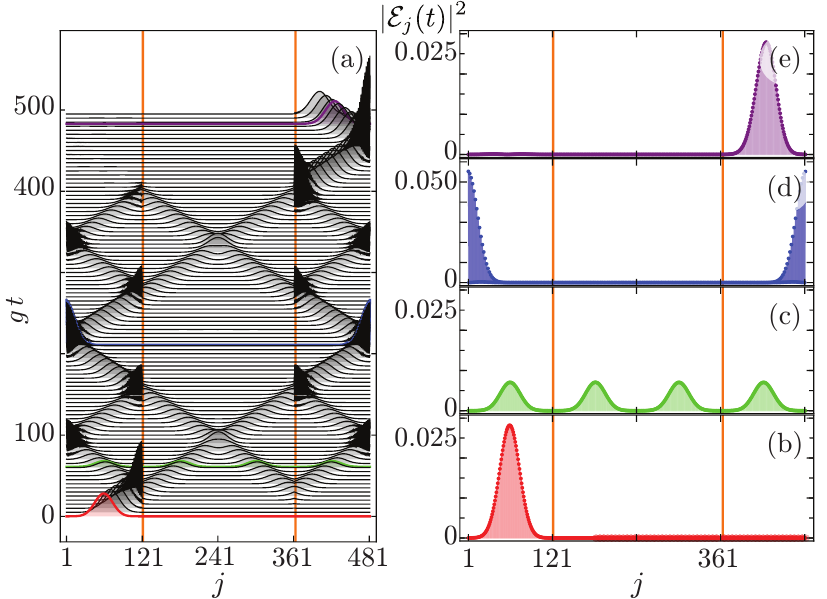}
	\caption{Non-local Mach-Zehnder interferometer. 
		(a) Squared field amplitude at the $j$-th resonator $\vert \mathcal{E}_{j}(t)\vert^{2}$ for different scaled evolution times:
		(b) $g t = 0$ before the wave packet interacts with the ring structure. 
		(c) $g t = 61$ when several wave packets propagate to the left and right. 
		(d) $g t = 211$ where two wave packets are reflecting at the ends of the CROW. 
		(e) $g t = 483$ where one wave packet propagates to the left. 
		All wave packets have waist $\sigma = 20$.}\label{fig:Fig7}
\end{figure*}

For an external resonator coupled to two sites of a finite size CROW, we place the external resonator such that there is the exact same number of 120 resonators from it to the endpoints of the CROW, and optimize the ring structure to contain twice that number, 240 resonators. 
In consequence, the effective optical path length in the ring structure is identical to twice the distance between the external resonator and the endpoints of the CROW creating a non-local Mach-Zehnder interferometer where a single input produces four optical paths, Fig. \ref{fig:Fig4}(a).
We optimize the zero resonant frequency detuning $\Delta \beta = 0$ and external coupling $g_{2} = 2 g$, producing transmittances $T_{1}=T_{2}=T_{3}=T_{4}=0.25$, for an initial wave packet, Fig. \ref{fig:Fig4}(b), to split into four almost identical wave packets, Fig. \ref{fig:Fig4}(c), two of them travel away from the external resonator and two travel inside the ring structure in opposite directions. 
All of them will meet again at the external resonator at the same time after those traveling away reflect from the endpoints of the CROW.
All of the secondary wave packets travel the same path length, and acquire the same phase.
They will interfere at the external resonator and recombine into two almost identical wave packets traveling away from the ring structure, Fig. \ref{fig:Fig4}(d).
These will reflect back and produce four wave packets, again, two of them travel away from the external resonator and two travel inside the ring structure in opposite directions.
These will recombine, again, at the external resonator and almost reproduce the initial wave packet profile traveling in the original direction, Fig. \ref{fig:Fig4}(e).

\section{Conclusions}\label{section:conclusions}

We studied propagation of Gaussian-like wave packets in coupled resonator optical waveguides, with constant envelope velocity and invariant waist, to demonstrate control of their scattering via a non-locally coupled external resonator, keeping velocity and waist invariant. 
The single-mode resonant frequency detuning and external resonator coupling, normalized with respect to the inter-site coupling strength, control the transmittances of these Gaussian-like wave packets.

For the sake of completeness, we calculate the reflectance and transmittance for an external resonator coupled to a single CROW site. 
This setup is the photonic simulation of a two-level atom coupled to a waveguide under the long wavelength approximation for a single excitation in the system. 
Numerical experiments are in good agreement with analytical results derived following standard waveguide-QED techniques.
We engineer a Mach-Zehnder interferometer using this configuration with an external resonator coupled to the middle site of a finite CROW to test the simplest example of multiple path interference in the model.

We explore non-local coupling using a single external resonator coupled to two sites in the CROW. 
A photonic simulation of a giant two-level atom non-locally coupled to a waveguide for a single excitation in the system.
Again, we calculate the analytical transmittances using standard waveguide-QED techniques.
They are in good agreement with our numerical experiments. 
We engineer three configurations to explore interference effects in this system focusing on the number of resonators in the CROW ring structure between the two coupling sites compared to the width of the initial Gaussian-like wave packet.
First, for a structure with more resonators than the initial wave packet width, we show effective exponential decay, via a series of plateaus and falls, of the total squared field amplitude trapped in the structure.
Second, for structures with a number of resonators of the order of the initial wave packet width, we show wave packet splitting; in particular, we engineer a configuration that produces two pairs of identical Gaussian-like wave packets that propagate away from the external resonator with the same waist than the initial wave-packet.
Third, we place the external resonator in a finite CROW such that the structure contains twice the number of sites than those from the external resonator to the endpoints of a finite CROW. 
Thus, we are able to produce a single input interferometer producing four optical paths of equal length that, after four interactions with the external resonator, reconstruct the initial Gaussian-like wave packet.

We look forward to studying more complex scattering scenarios where a giant atom, played by an external resonator, couples in a structured manner with the modes in the continuum, played by the CROW \cite{Jaramillo2020}, or study the effect of multiple external resonators coupled in varied configurations to the CROW where their single-mode resonant frequency is controlled in time.

\begin{acknowledgments}
F.~H.~M.-V. and B.~J.-A. acknowledge financial support from Catedras CONACYT Fellowship Program 551.
\end{acknowledgments}

\appendix
\section{Analytical calculations}\label{appendix:scattering}

An infinite CROW shows invariance under translations in the system allowing traveling signals \cite{Szameit2008}.
Introducing an external resonator breaks this symmetry producing reflected and transmitted signals.
This physical process can be described more accurately as follows: An infinite CROW array with resonator elements labeled as $j \in \mathbb{Z}$ has $N$ sites coupled to an external resonator. The coupled elements in the CROW are denoted by the set $\{p_{1}, p_{2},...,p_{N}\}$. 
Let $e^{ij\phi}$ denote an excitation arriving from the left, i.e. $j < \text{min}\left(p_{1}, p_{2},...,p_{N}\right)$, with wave number $\phi$. After reaching the coupling region in the array it will be partially reflected and transmitted. At each coupling site $p_{k}$ the signal splits into a backward and forward channel. Each splitted amplitude is represented by $\tau_{m}$ with $m \in \{1,2,\ldots,2N\}$. One can convince oneself that there will be $2N$ such amplitudes. The real number $|\tau_{m}|^{2}$ quantifies the transmittance of the signal in each channel. These fractions satisfy the constraint $\sum_{m}\vert\tau_{m}\vert^{2}=1$. In general $\tau_{m}=\tau_{m}(\phi)$, i.e. the amplitude depends on the wave number of the incident signal. If the initial pulse in the CROW is a superposition of multiple plane waves $\mathcal{E}_{\phi}(0)=\sum_{j=-\infty}^{\infty}\mathcal{E}_{j}(0)e^{ij\phi}$, each term in the sum will be transmitted with amplitude $\tau_{m}(\phi)$. Therefore, effective transmittance $T_{m}$ is given by the following expression,
\begin{align}\label{eq:app_full}
	T_{m} = 
	\frac{
		\int_{-\pi}^{\pi}\vert \tau_{m}(\phi)\mathcal{E}_{\phi}(0)\vert^{2} \mathrm{d}\phi
	}{
		\int_{-\pi}^{\pi}\vert\mathcal{E}_{\phi}(0)\vert^{2} \mathrm{d}\phi
	},
\end{align}
where the relation $\sum_{m}T_{m}=1$ still holds. 
Below, we discuss this process when an external resonator is coupled to one or two sites in the CROW.

\subsection*{One site coupling}

As discussed above, to compute the transmittance coefficients we need the amplitudes $\tau_{m}$, which are configuration dependent. 
First, we find these amplitudes for the case of one site in the CROW coupled to an external resonator, at site $p_{1}=k$. Here there are two amplitudes, $\tau_{1}$ and $\tau_{2}$. 
The excitation in the CROW and external resonator can be written as the following expressions,
\begin{equation}
	\begin{aligned}
		\mathcal{E}_{j}(t)&=
		\begin{cases}
			\left[e^{i(j-k)\phi}+\tau_{1} e^{-i(j-k)\phi}\right]e^{-i\omega t}	&j<k, \\
			\tau_{2} e^{i(j-k)\phi} e^{-i\omega t}								&j>k, \\
			\tilde{\mathcal{E}}_{k} e^{-i\omega t}								&j=k.
		\end{cases}\\
		\mathcal{E}_{e}(t)&=\tilde{\mathcal{E}}_{e}e^{-i\omega t},
	\end{aligned}
\end{equation}
where $	\tilde{\mathcal{E}}_{k}$ and $\tilde{\mathcal{E}}_{e}$ are the amplitudes in the $k$-th element of the CROW and the external resonator, in that order. 
Replacing the latter relations into the coupled mode equations, Eq. (\ref{eq:differenceq}), we find that indices $j<k-2$, produce the constraint $\omega=2g\cos\phi$. 
From the four coupled equations, including the external resonator, we solve for $\tau_{1}$ and $\tau_{2}$, 
\begin{align}\label{eq:app1}
	\tau_{1} = \frac{
		g_{2}^{2}
	}{
		g_{2}^{2}+2i g \sin \phi (2 g \cos \phi-\Delta\beta)
	},	
	\qquad 
	\tau_{2} = -\frac{
		2g \Delta\beta -4 g^{2} \cos \phi
	}{
		2g^{2} \Delta\beta-4g^{2} \cos \phi+i g_{2}^{2}\csc \phi
	}.
\end{align}
Using the latter relations we obtain the reflectance and transmittance coefficients, 
\begin{equation}
	\begin{aligned}
		&R = \int_{-\pi}^{\pi}\left| 
		\frac{
			g_{2}^{2}
		}{
			g_{2}^{2}+2i g \sin\phi (2 g \cos \phi-\Delta\beta)
		}\right|^{2}
		\vert\mathcal{E}_{\phi}(0)\vert^{2}\mathrm{d}\phi
		\bigg/
		\int_{-\pi}^{\pi}\vert\mathcal{E}_{\phi}(0)\vert^{2} \mathrm{d}\phi, \\ 
		&T = \int_{-\pi}^{\pi}\left| 
		\frac{
			2g \Delta\beta -4 g^{2}\cos \phi
		}{
			2g^{2} \Delta\beta-4g^{2}\cos \phi+i g_{2}^{2}\csc\phi
		}\right|^{2}
		\vert\mathcal{E}_{\phi}(0)\vert^{2}\mathrm{d}\phi
		\bigg/
		\int_{-\pi}^{\pi}\vert\mathcal{E}_{\phi}(0)\vert^{2} \mathrm{d}\phi.
	\end{aligned}
\end{equation}
Using the Gaussian profiles, $\mathcal{E}_{\phi}(0)= \sigma e^{-\frac{1}{2} \theta ^2 \sigma ^2-\theta \sigma ^2 \phi -\frac{1}{2} \phi \left(\sigma ^2 \phi +2 i j_{0}\right)}$ in the initial wave packet configuration, we evaluate the reflectance and transmittance coefficients in Section \ref{section:local}. 
When the single-mode resonant frequency detuning is not close to zero, we can approximate the integrals above to obtain Eq. \ref{eq:rtanalytic}.

\subsection*{Two site coupling}

Taking advantage of the insight from the previous case, we now turn to the case where the CROW is coupled to an external resonator at points $p_{1}=k_{1}$ and $p_{2}=k_{2}$. We have four amplitudes $\tau_{m}$ with $m=1,2,3,4$ as depicted in Fig. \ref{fig:Fig4}(a). In this process, an excitation arrives from the left $j<k_{1}$ and is scattered at $j=k_{1}$, where the first coupling site is located. A fraction of the transmitted signal will move to the ring and the remaining will be scattered at site $j=k_{2}$. Mathematically, this can be described as,
\begin{equation}
	\begin{aligned}
		\mathcal{E}_{j}(t)&=
		\begin{cases}
			\left[e^{i(j-k_{1})\phi}+\tau_{2} e^{-i(j-k_{1})\phi}\right] e^{-i\omega t}	&j<k_{1}, \\
			\tau_{1} e^{i(j-k_{1})\phi} e^{-i\omega t}									&j>k_{1}\text{ and } j \ll k_{2}, \\
			\tau_{3} e^{-i(j-k_{2})\phi} e^{-i\omega t}									&j \gg k_{1}\text{ and }j<k_{2} \\
			\tau_{4} e^{i(j-k_{2})\phi} e^{-i\omega t}									&j>k_{2}, \\
			\tilde{\mathcal{E}}_{k_{1}} e^{-i\omega t}									&j=k_{1},\\
			\tilde{\mathcal{E}}_{k_{2}} e^{-i\omega t}									&j=k_{2}
		\end{cases}\\
		\mathcal{E}_{e}(t)&=\tilde{\mathcal{E}}_{e}e^{-i\omega t},
	\end{aligned}
\end{equation}
where $\tilde{\mathcal{E}}_{k_1}$, $\tilde{\mathcal{E}}_{k_2}$ and $\tilde{\mathcal{E}}_{e}$ are the amplitudes in the sites $k_{1}$ and $k_{2}$ in the CROW and the external resonator, respectively. Again, after replacing this ansatz into the coupled mode equations, Eq. (\ref{eq:differenceq}), and taking into account the validity domains, we obtain the following relations,
\begin{equation}
	\begin{aligned}\label{eq:app2}
		\tau_{1} &=1-\frac{g_{2}^{2}}{2 \left[g_{2}^{2}+ i g \sin \phi (2 g\cos \phi-\Delta \beta )\right]}, \\
		\tau_{2} = \tau_{3} = \tau_{4} &= -\frac{g_{2}^{2}}{2 \left[g_{2}^{2}+ i g \sin \phi (2 g\cos \phi-\Delta \beta )\right]}.
	\end{aligned}
\end{equation}
From these relations, we see that $\vert\tau_{2}\vert^{2}=\vert\tau_{3}\vert^{2}=\vert\tau_{4}\vert^{2}$ and we can check $\sum_{m=1}^{4}\vert\tau_{m}\vert^{2}=1$.
They lead to the four transmittance coefficients,
\begin{equation}
	\begin{aligned}
		T _{1} &= \int_{-\pi}^{\pi}
		\left|
		1-\frac{
			g_{2}^{2}
		}{
			2 \left[g_{2}^{2}+ ig \sin \phi (2 g \cos \phi-\Delta \beta )\right]
		}
		\right|^{2}\vert\mathcal{E}_{\phi}(0)\vert^{2}d\phi
		\bigg/
		\int_{-\pi}^{\pi}\vert\mathcal{E}_{\phi}(0)\vert^{2} \mathrm{d}\phi, \\ 
		T_{2}=T_{3}=T_{4} &= \int_{-\pi}^{\pi} 
		\left|
		\frac{
			g_{2}^{2}
		}{
			2 \left[g_{2}^{2}+ i g \sin \phi (2 g\cos \phi-\Delta \beta )\right]
		}
		\right|^{2}\vert\mathcal{E}_{\phi}(0)\vert^{2}d\phi
		\bigg/
		\int_{-\pi}^{\pi}\vert\mathcal{E}_{\phi}(0)\vert^{2} \mathrm{d}\phi.
	\end{aligned}
\end{equation}
Much like in the previous case, using the Gaussian profiles we can evaluate the transmittance coefficients in Section \ref{section:nonlocal}. 
Again, when the resonant frequency detuning between the external resonator and CROW is not close to zero, we can approximate these integrals to obtain Eq. (\ref{eq:rtanalytict1234}).

%

\end{document}